\documentclass{article}

\usepackage{arxiv}

\usepackage[utf8]{inputenc} 
\usepackage[T1]{fontenc}    
\usepackage{hyperref}       
\usepackage{url}            
\usepackage{booktabs}       
\usepackage{amsfonts}       
\usepackage{nicefrac}       
\usepackage{microtype}      
\usepackage{lipsum}
\usepackage{graphicx}
\graphicspath{ {./images/} }
\usepackage{url}
\usepackage{amsmath}
\usepackage{amsfonts}
\usepackage{amssymb}
\usepackage{graphicx}
\usepackage{pgf,tikz}
\usepackage{pgfplots,subfig}
\usepackage{enumitem} 
\usetikzlibrary{shapes.multipart}
\definecolor{mycolor1}{rgb}{0.00000,0.44700,0.74100}%
\definecolor{mycolor2}{rgb}{0.85000,0.32500,0.09800}%
\pgfplotsset{ 
  compat=newest, 
   legend style =
  {font=\footnotesize \sffamily},
  label style = {font=\small\sffamily},
every tick label/.append style={font=\small}
  }

\renewcommand{\eqref}[1]{Eq.~(\ref{#1})}  

\title{A multi-layer network model to assess school opening policies during the COVID-19 vaccination campaign}

\author{
 Christian Bongiorno \\
  Université Paris-Saclay, CentraleSupélec,\\
  Mathématiques et Informatique pour la Complexité et les Systèmes,\\
  91190, Gif-sur-Yvette, France.
\\
  \texttt{christian.bongiorno@centralesupelec.fr} \\
   \And
 Lorenzo Zino \\
  Faculty of Science and Engineering,\\ University of Groningnen, 9747 AG Groningen.\\
  \texttt{lorenzo.zino@rug.nl}
}

\begin{document}
\maketitle
\begin{abstract}
We propose a multi-layer network model for the spread of COVID-19 that accounts for interactions within the family, between schoolmates, and casual contacts in the population. We utilize the proposed model ---calibrated on epidemiological and demographic data--- to investigate current questions concerning the implementation of non-pharmaceutical interventions (NPIs) during the vaccination campaign. Specifically, we consider scenarios in which the most fragile population has already received the vaccine, and we focus our analysis on the role of schools as drivers of the contagions and on the implementation of targeted intervention policies oriented to children and their families. We perform our analysis by means of a campaign of Monte Carlo simulations. Our findings suggest that, in a phase with NPIs enacted but in-person education, children play a key role in the spreading of COVID-19. Interestingly, we show that children's testing might be an important tool to flatten the epidemic curve, in particular when combined with enacting temporary online education for classes in which infected students are detected. Finally, we test a vaccination strategy that prioritizes the members of large families and we demonstrate its good performance. We believe that our modeling framework and our findings could be of help for public health authorities for planning their current and future interventions, as well as to increase preparedness for future epidemic outbreaks.
\end{abstract}

\section{Introduction}\label{sec:introduction}

The ongoing COVID-19 pandemic has called for an unprecedented mobilization of the scientific community toward understanding the transmission mechanisms, designing and assessing non-pharmaceutical intervention policies (NPIs) to mitigate its spread, and developing effective vaccines. Within this joint effort, a forefront role has been played by the development of accurate mathematical models to predict the spread of the pandemic and assess the effectiveness of the implementation of different NPIs, such as wearing face masks, enforcing social distancing, and enacting travel bans~\cite{estrada2020covid,Vespignani2020model,Bertozzi2020}. These works, grounded in the theory of mathematical modeling of epidemic diseases~\cite{Bailey1975,Hethcote2000} and on its recent application on complex networks~\cite{RevModPhys.87.925,Mei2017,Nowzari2016,Pare2020review}, have provided effective tools to assist public health authorities, by highlighting the epidemic risk, predicting its spatial and temporal spread, indicating limitations of the NPIs currently enacted, and suggesting potential strategies to improve them~\cite{chinazzi2020effect,Giordano2020,gatto2020,dellarossa2020,parino2021,arenas2020,kohler2020robust,carli2020}. 

From the inception of the epidemic outbreak, pharmaceutical researchers have started working at an unprecedented pace toward developing vaccines for the novel coronavirus SARS-CoV-2 ---which is the virus responsible for the COVID-19 disease--- achieving the astonishing goal of developing, testing, and obtaining approval by national regulatory authorities for several vaccines in less than one year\cite{owd_vaccination}. Hence, as of February 2021, many countries around the world are facing an unprecedented vaccination campaign~\cite{owd_vaccination}, while struggling with a second or third wave of the epidemic outbreak~\cite{whoSituation}. Even in this phase, mathematical models are valuable supports to assist public health authorities in their decisions on the vaccination strategies and on the NPIs that should be implemented during these phases~\cite{Grauer2020,Bubar2021,Truszkowska2021,foy2021}. 

In this work, we focus on the next phases of the vaccination campaigns, that is, when the most fragile part of the population ---those at higher risk of developing serious complications--- is already vaccinated; however, the large majority of the population ---including most of the workers and school children--- is still susceptible to the disease. Massive epidemic outbreaks are still possible during these phases and, thus, NPIs are still key to avoid the resurgence of the epidemic disease. Among the questions arising on the optimal calibration of NPIs, the policies concerning the management of schools and children play a crucial role for their impact on the education system and on their families. Moreover, the impossibility of reducing contagions in schools by means of vaccinations ---trials for vaccines on children started just as of February 2021~\cite{nyt}--- makes crucial to understand how to calibrate NPIs in order to flatten the epidemic curve. For these reasons, understanding the effect of different policies for school opening, including increasing the testing rate for children and implementing temporary online education in order to home-isolate the entire class whenever a child is tested positive in that class, is a problem of paramount importance.

In this vein, we propose a temporal network model~\cite{Holme2012,Holme2015} with a multi-layer structure~\cite{Kivela2014}, tailored to capture the contagions between children in schools and in their families. Specifically, the proposed model is developed on three layers: a \emph{family layer}, which represents the interactions between family members living in the same household; a \emph{school layer} that models the interactions between children in the same class; and a time-varying \emph{contact layer}, which captures casual interactions between adults, for instance, in shops or in public transport. While the family layer is assumed to be fixed, the other two layers are time-varying, allowing to capture different phenomena. The school layer is formed by a backbone of time-invariant edges ---modeling the classes--- which may be temporarily removed due to home-isolation. The time-varying contact layer, instead, is generated in a stochastic fashion, representing casual contact between adults. Specifically, we adopt an activity-driven network model~\cite{Perra2012,Zino2016}, which has emerged as a valuable modeling framework to generate heterogeneous time-varying networks of interactions.

We combine the proposed network model with a disease progression mechanism tailored to COVID-19. Specifically, we consider an extension of a stochastic susceptible--exposed--infectious--removed (SEIR) model on networks~\cite{RevModPhys.87.925,Zino2017}, in which further compartments are added to account for vaccinations and for the presence of asymptomatic unaware infectious individuals, and in which heterogeneity between children and adults is incorporated by including two different classes of individuals. The model is calibrated by generating a population consistent with the US Census demographic data~\cite{nson,nces_report}, and by setting the epidemic parameters of COVID-19 following reliable estimations from the epidemiology literature~\cite{Prem2020,Billah2020R0}.

We utilize our model to investigate the role of children and schools in the spreading of COVID-19 and to assess the effectiveness of different policies during the current and future stages of the vaccination campaign. Our analyses, performed through an extensive campaign of Monte Carlo simulations, allow us to draw some conclusions. First, we use the model to support the intuition that children play a key role in the spreading of COVID-19, and thus ---being the vaccination of children still not viable~\cite{nyt}--- the management of NPIs in schools seems crucial to keep the infections under control while vaccinating the adults. Second, massive testing campaigns in schools seem to be effective in mitigating the spread. However, these campaigns may be practically unfeasible, since they may require detecting at least $70\%$ of the infections (including asymptomatic) to be able to flatten the curve ---an objective that might be far beyond the current estimates~\cite{Pullano2020}. Third, the enforcing of online education for schoolmates of detected infected children seems to be an effective practice to keep the number of infections under control, in combination with a moderate testing campaign in schools. Finally, we find that prioritizing vaccination of large families may be a valuable strategy to reach herd immunity faster, limiting the need for massive testing campaigns in schools. We believe that these findings might be of help to assist public health authorities during these crucial phases of the fight against COVID-19. Moreover, the generality of our modeling framework suggests that it could be a valuable tool to investigate issues that may arise in the future stages of the pandemic, and to increase preparedness for future epidemic outbreaks.

The rest of the paper is organized as follows. In Section~\ref{sec:model}, we present our multi-layer network epidemic model. In Section~\ref{sec:calibration}, we calibrate our general modeling framework to investigate the ongoing COVID-19 challenges. In Section~\ref{sec:results}, we present our main results and we discuss their implications. Section~\ref{sec:conclusions} concludes the paper by summarizing the take-home message and outlining future research directions.

\section{Model}\label{sec:model}

\subsection{Population}

We consider a population of $n$ individuals, $n\in\mathbb Z^+$, indexed by $\mathcal V=\{1,\dots,n\}$, where $\mathbb Z^+$ is the set of non-negative integer numbers. Individuals are divided into two types: \emph{children} $\mathcal C=\{1,\dots,\tilde n\}$ and \emph{adults} $\mathcal A=\{\tilde n+1,\dots,n\}$. The entire population is partitioned into a set $\mathcal F=\{\mathcal F_1,\dots\mathcal F_k\}$ of $k$ mutually exclusive families (so that $\mathcal V=\bigcup_{\ell=1}^k F_\ell$). Similarly, children are partitioned into a set $\mathcal S=\{\mathcal S_1,\dots\mathcal S_m\}$ of $m$ mutually exclusive school classes (so that $\mathcal C=\bigcup_{\ell=1}^m S_\ell$). We assume that the population and its partitioning in families and school classes remain constant throughout the duration of the epidemic outbreak.

\begin{figure}
    \centering
    \includegraphics{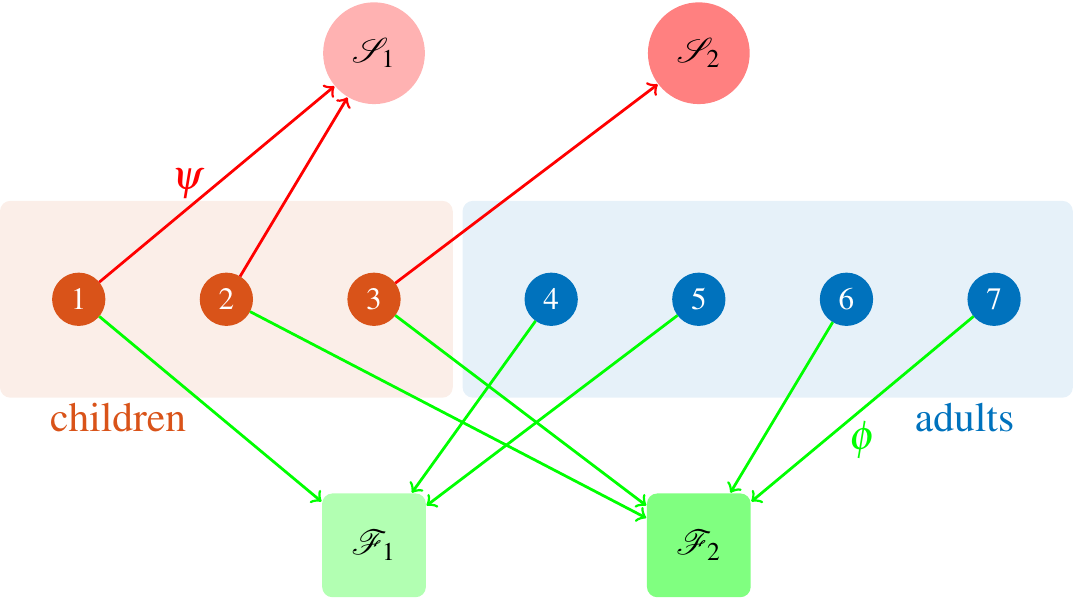}
    \caption{Schematic of the population structure. In this simple example, three children $\mathcal C=\{1,2,3\}$ and four adults $\mathcal A=\{4,5,6,7\}$ are partitioned into two school classes (red circles above) and two families (green squares below) by means of the functions $\psi$ and $\phi$, respectively.}
    \label{fig:population}
\end{figure}

We define two functions $\phi$ and $\psi$ that associate each individual with his or her family, and each child with his or her class. Specifically, we denote by $\phi:\mathcal V\to\mathcal F$ the function that associates each individual with the corresponding family and by $\psi:\mathcal C\to\mathcal S$ the function that associates each child with his or her class. Hence, each adult $i\in\mathcal A$ is associated with the family $\phi(i)$, while each child $j\in\mathcal C$ is associated with the family $\phi(j)$ and the class $\psi(j)$. A schematic of the population structure is illustrated in Fig.~\ref{fig:population}.

\subsection{Network}

In this section, we define a time-varying multi-layer network structure~\cite{Kivela2014}, which accounts for the presence of steady interactions within each family, intermittent interactions in schools, and casual time-varying contacts~\cite{Holme2012,Holme2015}. To this aim, we propose a three-layered undirected network structure $\mathcal G=({\mathcal V,\mathcal E_{F},\mathcal E_S(t),\mathcal E_C(t)})$, $t\in\mathbb Z^+$, in which the three layers capture the interactions between family members living in the same household, between schoolmates, and other casual contacts, respectively. Note that $\mathcal{E}_F$ is assumed to be constant, while $\mathcal{E}_{S}(t)$ and $\mathcal{E}_{C}(t)$ are time-varying, reflecting the assumption that ---at least on a time-scale relative to an epidemic outbreak--- families do not change, while the implementation of online education may reduce the physical interactions between classmates, and casual contacts between adults may vary day-to-day. The three layers are defined as follows.

The \textit{family layer} is defined by the time-invariant edge set $\mathcal E_{F}\subseteq\mathcal V\times\mathcal V$, which captures the interactions between family members, that is,
\begin{equation}
    (i,j)\in\mathcal E_F\iff \phi(i)=\phi(j)\,.
\end{equation}
Hence, the family layer is formed by a set of cliques, connecting all the individuals in the same family $\mathcal F_\ell$. 

The \emph{school layer} is defined by an (intermittent) time-varying edge set $\mathcal E_{S}(t)$, which represents the interactions between schoolmates at time $t$. Specifically, we define the \emph{school backbone} $\bar{\mathcal E}_S\subseteq\mathcal C\times\mathcal C$ as
\begin{equation}
    (i,j)\in\bar{\mathcal E}_S\iff i,j\in\mathcal C\quad\text{and}\quad\psi(i)=\psi(j)\,,
\end{equation}
that is, two children are connected by a link on the school backbone if and only if they are in the same class. Hence, the school backbone forms a set of cliques, connecting all the children in the same set $\mathcal S_\ell$. The school layer at time $t$ is a subset of such a backbone, that is, $\mathcal E_S(t)\subseteq \bar{\mathcal E}_S$, accounting for periodic school closures (for instance, during weekends) and for the implementation of NPIs (for instance, by enforcing online education for infected children, or for their entire class). To this aim, we define a variable $A_i(t)\in\{0,1\}$, $i\in\mathcal C$, termed \emph{home-isolation state} representing whether $i$ goes to school at time $t$ ($A_i(t)=1$) or is home-isolated ($A_i(t)=0$). Finally, we define the school layer at time $t$ as follows:
\begin{equation}
    (i,j)\in\mathcal E_S(t)\iff (i,j)\in\bar{\mathcal E_S}\quad\text{and}\quad A_i(t)=A_j(t)=1\,.
\end{equation}

The \textit{contact layer} is defined by the time-varying edge set $\mathcal E_C(t)\subseteq\mathcal A\times\mathcal A$, representing the casual interactions that are generated, for instance, in a shop, or in public transportation, or at work. We assume that only adults individuals are involved in this type of interaction, while children only interact with their family members and their classmates. This is consistent with the presence of mobility restrictions and with the closure of most non-essential activities during the early vaccination stages.

\begin{figure}
    \centering
    \includegraphics[width=\textwidth]{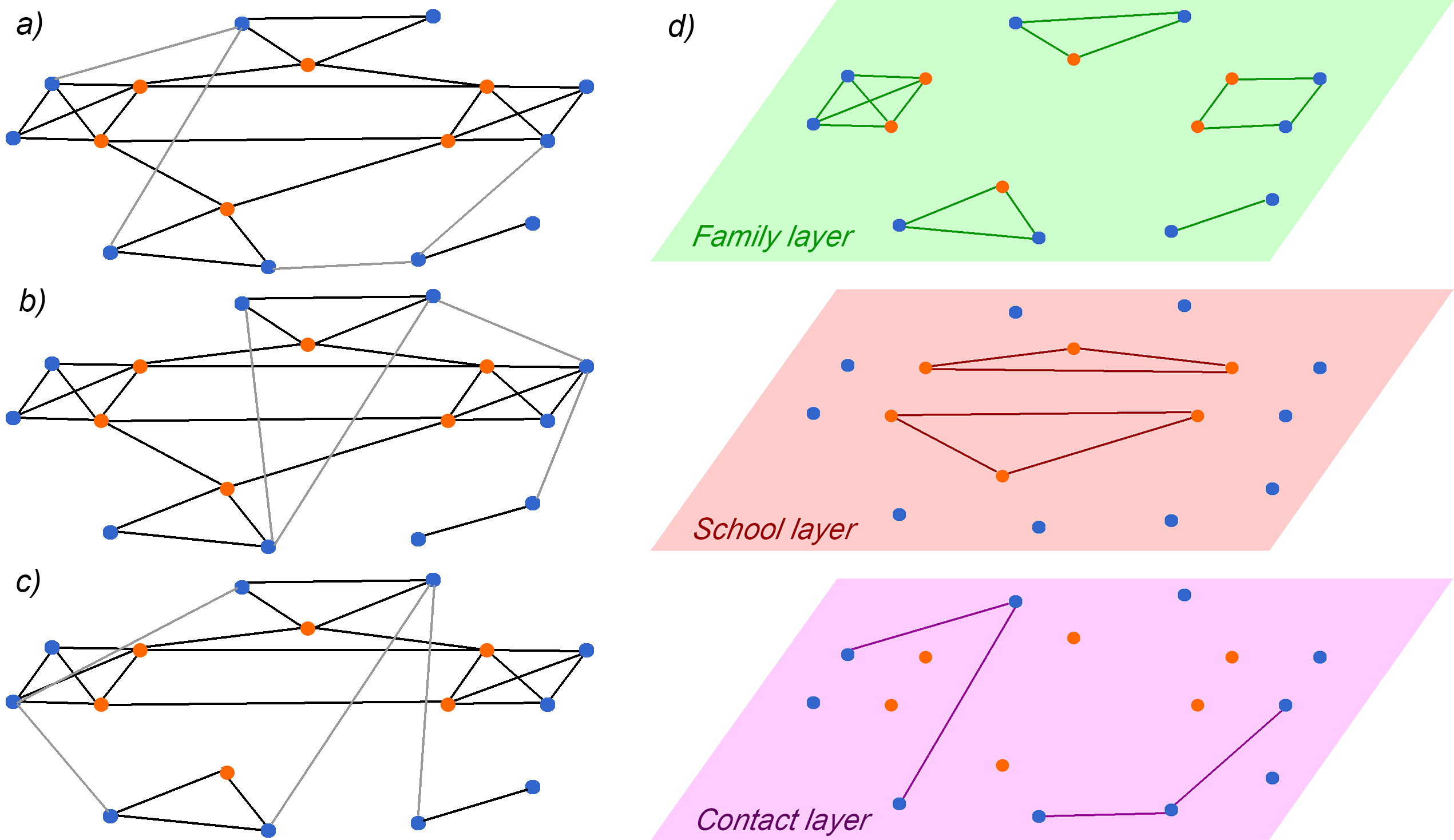}
    \caption{Schematic of the network structure. In a)--c), we illustrate three consecutive time instances of the network; the links in gray correspond to the temporal links generated on the contact layer. Orange nodes represent children, blue nodes represent adults. Note that in c), one of the children is home-isolated (the one below). Hence, all the links with school-mates on the school layer are temporarily removed. In d), we show the three layers and the corresponding links at the time instant illustrated in a).}
    \label{fig:network}
\end{figure}

We generate the contact layer by using an extension of a discrete-time activity-driven network (ADN)~\cite{Perra2012}. We adopt ADNs ---which have emerged as a powerful modeling framework to realistically reproduce time-varying heterogeneous networks~\cite{Starnini2013}--- because of their flexibility~\cite{RizzoPRE2014,PhysRevLett.114.108701, Nadini2018,Bongiorno2019} which allows us to incorporate the specific features of our model such as different home-isolation policies, and for their amenability to efficiently perform fast numerical simulations~\cite{Rizzo2016Ebola}. Similar to a standard ADN~\cite{Perra2012}, each adult is characterized by a constant parameter $a_i\in[0,1]$, $i\in\mathcal A$, called \emph{activity}, expressing his or her propensity to interact with others, and all the adults have a common parameter $m$ that represents the number of interactions that active individuals initiate. Moreover, similar to children, also each adult is associated with a \emph{home-isolation state} $A_i(t)\in\{0,1\}$, $i\in\mathcal A$, representing whether $i$ is allowed to have social interactions at time $t$ ($A_i(t)=1$), or if he or she is home-isolated ($A_i(t)=0$). The contact layer is generated according to the following algorithm:
\begin{enumerate}
    \item the time is initialized at $t=0$;
    \item the edge set is initialized as an empty edge set $\mathcal E_C(t)=\emptyset$;
    \item each adult individual $i\in\mathcal A$ that is not home isolated ($A_i(t)=1$) activates with probability equal to $a_i$, independent of the others and of the previous history of the process;
    \item if $i$ activates, then the individual generates $m$ undirected links with a $m$-tuple of (non-home-isolated) adults outside their family ($\phi(i)$), selected uniformly at random in the set $\{j:j\in \mathcal A, \phi(j)\neq \phi(i), A_j(t)=1\}$; the generated links are added to the set $\mathcal E_C(t)$;
    \item the time-step is increased by $1$, and the algorithm resumes from item 2).
\end{enumerate}
The multi-layer network structure obtained is illustrated in Fig.~\ref{fig:network}.

\subsection{Disease progression and spreading}

\begin{figure}
\centering
    \includegraphics{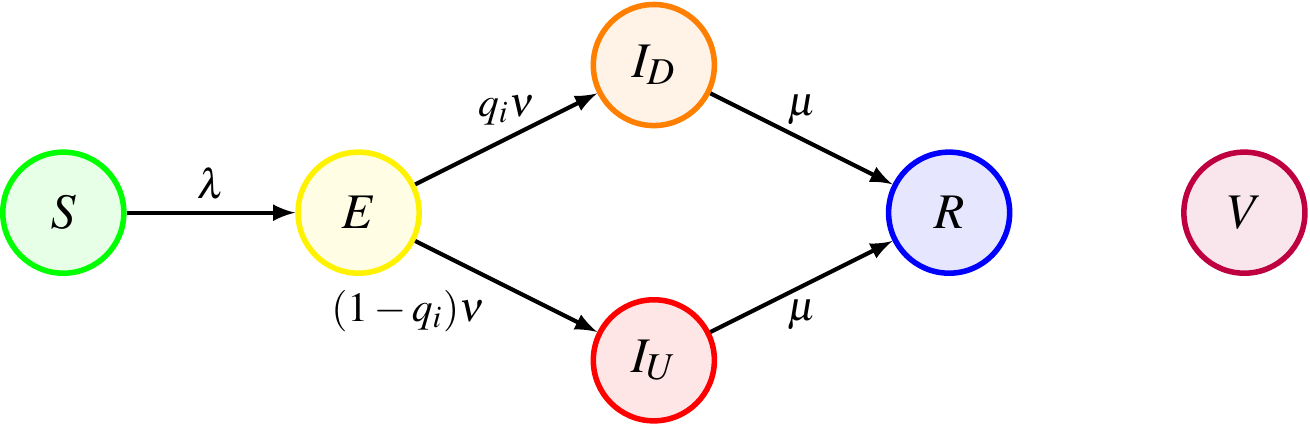}
\caption{State transitions characterizing the epidemic spreading model. Susceptible individuals ($S$) that have interactions with infectious individuals ($I_D$ and $I_U$) may become exposed ($E$), and then infectious. Infectious individuals can be either detected ($I_D$) or unaware ($I_U$). Then, they may recover or die, becoming removed ($R$). Vaccinated individuals cannot contract the disease. }\label{fig:schematic}
\end{figure}

We consider an extension of a stochastic network SEIR model on networks~\cite{RevModPhys.87.925,Zino2017}, in which we have included two additional compartments to account for vaccinations and for the presence of asymptomatic unaware infectious individuals. Specifically, at discrete time instant $t\in\mathbb Z^+$, each individual $i\in\mathcal V$ is characterized by a variable $X_i(t)\in\{S,E,I_D,I_U,R,V\}$, representing the  \emph{health state} of the individual at time $t$. The state $S$ represents \emph{susceptible} individuals, who are healthy and can be infected by the disease, while individuals that have been vaccinated and are immune to the disease are denoted by $V$. Susceptible individuals who have been \emph{exposed} to the disease and are thus infected, but not yet infectious, are represented by $E$. Infectious individuals are divided into two compartments: those that are detected (due to the emergence of symptoms or because of receiving a positive test), denoted by $I_D$, and those that are asymptomatic and not tested, and thus unaware ($I_U$). Finally, individuals that recover (or die) are denoted by $R$. We assume that recovered individuals become immune to the disease for the time-horizons considered in our analysis (in all our simulations, we focus on a time-window that does not exceed 6 months), as found in recent studies~\cite{Gudbjartsson2020immunity,Dan2021immunity}. The progression of the disease is described in the following and illustrated in Fig.~\ref{fig:schematic}.

At each time-step $t$, susceptible individuals ($X_i(t)=S$) may be exposed and thus infected with COVID-19. The \emph{per-contact infection probability} $\lambda\in[0,1]$, is a constant parameter that captures the probability that the disease is transmitted through a physical contact with an infectious individual ($I_D$ and $I_U$). 
Hence, the contagion probability for each individual $i\in\mathcal V$ is equal to
\begin{equation}\label{eq:EI}
    \mathbb P[X_i(t+1)=E|X_i(t)=S]=1-(1-\lambda)^{N^F_i(t)+N^S_i(t)+N^C_i(t)}\,,
\end{equation}
where, denoted by $|\cdot|$ the cardinality of a set, 
\begin{subequations}\label{eq:N}
\begin{align}
&N^F_i(t)=|\{j\in\mathcal V:(i,j)\in\mathcal E_F,\,X_j(t)\in\{I_U,I_D\}\}|\,,\\
&N^S_i(t)=|\{j\in\mathcal V:(i,j)\in\mathcal E_S(t),\,X_j(t)\in\{I_U,I_D\}\}\,|,\\
&N^C_i(t)=|\{j\in\mathcal V:(i,j)\in\mathcal E_C(t),\,X_j(t)\in\{I_U,I_D\}\}|\,,
\end{align}
\end{subequations}
are the number of infectious individuals that have a link with individual $i$ at time $t$ on the family, school, and contact layer, respectively. 

Besides the contagion, at each time-step $t$, each exposed individual $E$ may become infectious, with probability $\nu\in[0,1]$. Infectious individuals may be aware ($I_D$) due to being symptomatic or receiving a positive test, or unaware ($I_U$) due to being asymptomatic and untested. We denote by $q_i\in[0,1]$ the probability that individual $i$ is detected if infected. We assume that such a probability depends on whether the individual is a children or an adult, as a consequence of different probabilities of developing symptoms and different testing policies that can be implemented for the two types of individuals. Hence, we introduce two parameters $q_a\in[0,1]$ and $q_c\in[0,1]$ to model the \emph{adults detection rate} and the \emph{children detection rate}, respectively, and we set 
\begin{equation}
    q_i=\left\{\begin{array}{ll}
    q_c&\text{ if }i\in\mathcal C\,,\\
        q_a&\text{ if }i\in\mathcal A\,.
    \end{array}\right.
\end{equation}
Infectious individuals recover or die and become removed ($R$), with probability $\mu\in [0, 1]$. 
These transitions, illustrated in Fig.~\ref{fig:schematic}, are thus governed by the following probabilistic rules:
\begin{subequations}\label{eq:other_transitions}\begin{align}
&\mathbb P[X_i(t+1)=I_D\,|\,X_i(t)=E]=q_i\nu\,,\\
&\mathbb P[X_i(t+1)=I_U\,|\,X_i(t)=E]=(1-q_i)\nu\,,\\
&\mathbb P[X_i(t+1)=R\,|\,X_i(t)\in\{I_D,I_U\}]=\mu\,.
\end{align}\end{subequations}
Table~\ref{tab:notation} summarizes the notation use throughout this paper.

\begin{table}
\centering
\begin{tabular}{rl}
Notation & Meaning
\\ \hline
$\mathcal V=\{1,\dots,n\}$  & population\\ 
$\mathcal C=\{1,\dots,\tilde n\}$  & children\\ 
$\mathcal A=\{\tilde n+1,\dots,n\}$  & adults\\ 
$\mathcal F=\{\mathcal F_1,\dots,\mathcal F_k\}$  & families\\ 
$\mathcal S=\{\mathcal S_1,\dots,\mathcal S_m\}$  & school classes\\
$\phi:\mathcal V\to\mathcal F$  & function that associates individuals to their families\\
$\psi:\mathcal C\to\mathcal S$  & function that associates children to their classes\\
$\mathcal E_F\subseteq \mathcal V\times\mathcal V$  & family layer \\ 
$\bar{\mathcal E_S}\subseteq \mathcal C\times\mathcal C$  & school backbone edge set \\ 
$\mathcal E_S(t) \subseteq \bar{\mathcal E_S}$ & school layer at time $t$ \\ 
$\mathcal E_C(t) \subseteq \mathcal A\times\mathcal A$ & contact layer at time $t$\\ 
$a_i\in[0,1]$& activity of individual $i$\\
$m\in\mathbb Z^+$& interactions initiated by an active individual\\
$X_i(t)\in\{S,E,I_D,I_U,R,V\}$& health state of individual $i$ at time $t$\\
$A_i(t)\in\{0,1\}$& home-isolation state of individual $i$ at time $t$\\
$\lambda\in[0,1]$& per-contact infection probability\\
$\nu\in[0,1]$& probability of becoming infectious\\
$\mu\in[0,1]$& recovery probability\\
$q_a\in[0,1]$& adults detection rate\\
$q_c\in[0,1]$& children detection rate\\
\end{tabular}
\caption{Notation used in the paper.}
\label{tab:notation}
\end{table}

\subsection{Dynamics}

Under the reasonable assumption that whether an individual is home-isolated at time $t$ depends only on the health state of the system at that time and on the time instant $t$, that is, that $A_i(t)$ is a deterministic function of $X(t)$ and of $t$, the network formation process of the two time-varying layers at time $t$ is a function of $X(t)$ and of $t$. Hence, the stochastic process $X(t)$ is ultimately a Markov chain on the state space $\{S,E,I_D,I_U,R,V\}^n$, whose state transitions are governed by Eqs.~(\ref{eq:EI}) and (\ref{eq:other_transitions}). Furthermore, if $A_i(t)$ depends on $t$ only through $X(t)$, then the Markov chain is time-invariant~\cite{levin2006book}. The latter is the case in which home-isolation policies are feedback of the state (for instance, if detected individuals, their family members, and/or their classmates are home-isolated), but no time-dependent policy is enacted (for instance, school attendance on alternating days or weeks). The Markovianity of the process $X(t)$ allows performing fast simulations of the systems, to shed light on the role of children and schools in the transmission of COVID-19, and to investigate the effectiveness of different home-isolation and vaccination strategies, as illustrated in Section~\ref{sec:results}.

\section{Model calibration and simulation setting}\label{sec:calibration}

Before presenting the findings of our simulation studies, we provide some details on the model calibration, based on demographic and epidemiological data, and on the setting we have designed to perform the simulations.

\subsection{Population and network}

We generate a network composed of $k=3,000$ families, each family is composed by $n_a=2$ adults and a variable numbers of children $n_s$, generated from a Poisson random variable (r.v.) with an expected value equal to $\langle n_s \rangle = 1.93$, each one independent of the others. Such a variable is calibrated on the average number of school children for families in the US~\cite{nson}. Note that the total number of individuals in the network $n$ is a r.v. with an expected value equal to $\langle n \rangle=11790$, being equal to $6,000$ adults and the sum of $3,000$ independent and identically distributed Poisson r.v.s with an expected value equal to $1.93$.

The children are randomly associated with their classes. Specifically, the number of children in each class $n_c$ is selected at random, according to a Poisson distribution with an expected value equal to $\langle n_c \rangle=20.9$, consistent with the average class size in US public schools~\cite{nces_report}.

The contact layer is generated according to a discrete-time ADN~\cite{Perra2012}. Specifically, the activity of $i\in\mathcal A$ is a power-law distributed r.v. with lower-barrier at $a_{\min}=0.1$, upper-barrier at $a_{\max}=1$, and exponent equal to $\alpha=-2.09$, as in~\cite{Aiello2001}. The number of connections initiated by each active individuals $m$ is computed as follows. Based on empirical observations, the authors in~\cite{Mossong2008} identify that active adults  have on average $18$ daily interactions (including those within the family, and those initiated by other individuals). Hence, we set the value of $m$ such that the expected degree of active adults matches with this estimation, rounded to the closest integer. Specifically, we enforce
\begin{equation}
  18=(1+\langle a\rangle)m+\langle k_{f} \rangle\implies m = \text{round}\left(\frac{18 - \langle k_{f} \rangle }{1+\langle a \rangle}\right)=12\,,
\end{equation}
where
\begin{equation}
    \langle a \rangle = \frac{\int_{a_{\min}}^1  x^{1+\alpha} dx}{\int_{a_{\min}}^1  x^{\alpha} dx}= \frac{\int_{0.1}^1  x^{-1.09} dx}{\int_{0.1}^1  x^{-2.09} dx}=0.247
\end{equation}
is the average activity, and 
\begin{equation}
    \langle k_f \rangle =n_a+ \langle n_s \rangle -1 = 2.93
\end{equation} 
is the average number of in-family links. Finally the average degree during week-days is obtained as the weighted average of the degree of the two types of individuals on the three layer, that is,
\begin{equation}\label{eq:k}
    \langle k \rangle = \frac{n_a}{n_a+ \langle n_s \rangle }\left( 2 \langle a \rangle  m +  \langle k_f \rangle \right)+\frac{\langle n_s \rangle }{n_a+\langle n_s \rangle}\left( \langle k_f \rangle + \langle n_c\rangle -1  \right)  = 15.27\,.
\end{equation}

\subsection{COVID-19 epidemic parameters}

The epidemic parameters $\lambda$, $\nu$, and $\mu$ are set from epidemiological data as follows. Reliable estimations of the average latency period $\tau_E=6.4$ days and of the average period of communicability $\tau_I=5$ days are available~\cite{Prem2020}. From these data, similar to~\cite{Prem2020}, we obtain
\begin{subequations}\begin{align}
&\nu=1-\exp\left(-\frac{1}{\tau_E}\right)=0.1447\,,\\
&\mu=1-\exp\left(-\frac{1}{\tau_I}\right)=0.1813\,.
\end{align}\end{subequations}

The per-contact infection probability $\lambda$ is estimated from the basic reproduction number $R_0$, which is the average number of secondary infections generated by an infectious individual, assuming that the rest of the population is susceptible. Specifically, according to the results of a recent meta-analysis on the estimation of the basic reproduction number~\cite{Billah2020R0}, we set $R_0=2.87$. Hence, since $\tau_I$ is the average time that an individual is infectious, we write
\begin{equation}
    R_0=\lambda \langle k \rangle \tau_I\,,
\end{equation}
where the average degree $\langle k \rangle=  15.27 $ is computed in \eqref{eq:k}. Hence, we conclude that
\begin{equation}
 \lambda=\frac{R_0}{ \langle k \rangle \tau_I}=0.0376\,.
\end{equation}

The detection rates $q_a$ and $q_c$ are control parameters that reflect the effort and the effectiveness of testing practices and they are lower-bounded by the symptomatic rates $q_a\geq0.25$ and $q_c\geq 0.12$, as suggested in~\cite{edge_health} and~\cite{Han2021}, respectively.

\subsection{Home-isolation policies}\label{sec:home}

Our work considers the implementation of different home-isolation policies. Differently from vaccinations, which are set at the beginning of each simulation, isolation policies are dynamical. 
Common to all policies we have that individuals that are detected ($I_D$), are always home-isolated, that is,
\begin{equation}
    X_i(t)=I_D\implies A_i(t)=0.
\end{equation}
Besides this basic rule, further policies may be enacted to enforce home-isolation of non-detected individuals that may be (potentially) infected. Specifically, we consider two policies:
\begin{enumerate}[label=\Alph*.]
    \item the \emph{family-isolation} policy, for which when an individual is detected, then all his or her family members are home-isolated, that is
    \begin{equation}
    X_i(t)=I_D\implies A_j(t)=0,\qquad \forall j\in \phi(i)\,;
\end{equation}
    \item the \emph{class-isolation} policy, for which when an individual is detected, then all his or her schoolmates are home-isolated, that is
    \begin{equation}
    X_i(t)=I_D\implies A_j(t)=0,\qquad \forall j\in \psi(i)\,.
\end{equation}
\end{enumerate}

In all our simulations, we assume that the family-isolation rule is enacted, while in some of the simulations, we will also enforce the class-isolation policy. We will explicitly report when both policies are implemented.

\subsection{Vaccination strategies and initialization}\label{sec:vaccination}

In all our simulations, we initialize the system by setting a fraction $V\in[0,1]$ of the population in the vaccinated ($V$) health state. Such a fraction may vary across the simulation settings, as detailed in Section~\ref{sec:results} when describing the results. Fixed the fraction of vaccinated individuals, three different vaccination strategies are considered and compared:
\begin{enumerate}[label=\Roman*.]
    \item the \emph{uniform adult vaccination}, in which a fraction $V$ of the population chosen uniformly at random among the adults is vaccinated;
    \item the \emph{uniform vaccination}, in which the fraction $V$ is chosen uniformly at random among the entire population, including children. Even though this strategy is not realistically viable in the current stage of the vaccination campaign~\cite{nyt}, its implementation in our simulations allows us to better understand the impact of children in the spreading of COVID-19; and
    \item a \emph{targeted vaccination}, in which, similar to the uniform adult vaccination, a fraction $V$ of the population is chosen for vaccination. However, the individuals are not selected uniformly at random, but vaccinated in decreasing order of size of the corresponding families.
\end{enumerate}
Note that, for different vaccination policies, the population eligible for vaccination is different. In particular, while for I and III the eligible population coincides with all the adults, for II the entire population is eligible to be vaccinated.

The rest of the population, that is, those not vaccinated, are initialized by randomly assigning $1\%$ of the population to the exposed health state ($E$), while all the others are susceptible ($S$). The parameters common to all the simulations are reported in Table~\ref{tab:parameters}.

\begin{table}
\centering
\begin{tabular}{lll}
 & Meaning &Value
\\ \hline
$k$& number of families&$3,000$\\
$n_a$& adults in each family&$2$\\
$n_s$& children in each family&Poisson r.v., mean $1.93$\\
$n_c$& children in each class&Poisson r.v., mean $20.9$\\
$a_i$& activity of node $i$&$[0.1,1]$, power law r.v., exponent $-2.09$\\
$m$& interactions of active individuals &12\\
$\lambda$& per-contact infection probability&0.0376\\
$\nu$& probability of becoming infectious&0.1447\\
$\mu$& recovery probability&0.1813\\
$q_a$& adults detection rate&--\\
$q_c$& children detection rate &--\\
$V$& population vaccinated &--\\
\end{tabular}
\caption{Value of the parameters used in the simulations. The last three parameters, namely $q_a$, $q_c$, and $V$, vary across the simulations and their values are explicitly reported when presenting the results. }
\label{tab:parameters}
\end{table}

\section{Results}\label{sec:results}

In this section, we utilize the model calibrated to COVID-19 to investigate several scenarios. Specifically, we will devote the first set of experiments toward shedding light on the role of children in the spreading of COVID-19. Then, the second set of simulations is performed to explore the effectiveness of increasing the testing of adults and children, showing that massive testing campaigns among children are necessary to effectively reduce the contagions. Then, we assess the performance of the \emph{class-isolation} policy presented in Section~\ref{sec:home}, demonstrating that it seems an effective strategy to flatten the epidemic curve without the need of massive (and often unfeasible) testing campaigns. Finally, we compare the uniform adult vaccination strategy with the prioritization of large families proposed in Section~\ref{sec:vaccination}, showing that the proposed strategy might be a viable strategy to reach herd immunity faster.

\subsection{Key role of children in the spreading of COVID-19}

In the first set of experiments, we investigate the role of children in the spreading of COVID-19. To remove any other confounding elements, we set the detection rates at their minimal values, coinciding with the symptomatic rates, that is, $q_a = 0.25$ and $q_c=0.12$, and we consider the simplest home-isolation policy A, described in Section~\ref{sec:home}, in which only family members of detected individuals are enforced to home-isolate themselves. To highlight the role of children in the spreading process, we report the number of infections among children and adults separately.

\begin{figure}
    \centering
        \subfloat[Temporal evolution of the outbreak]{\label{fig:inf}\includegraphics[width=7cm]{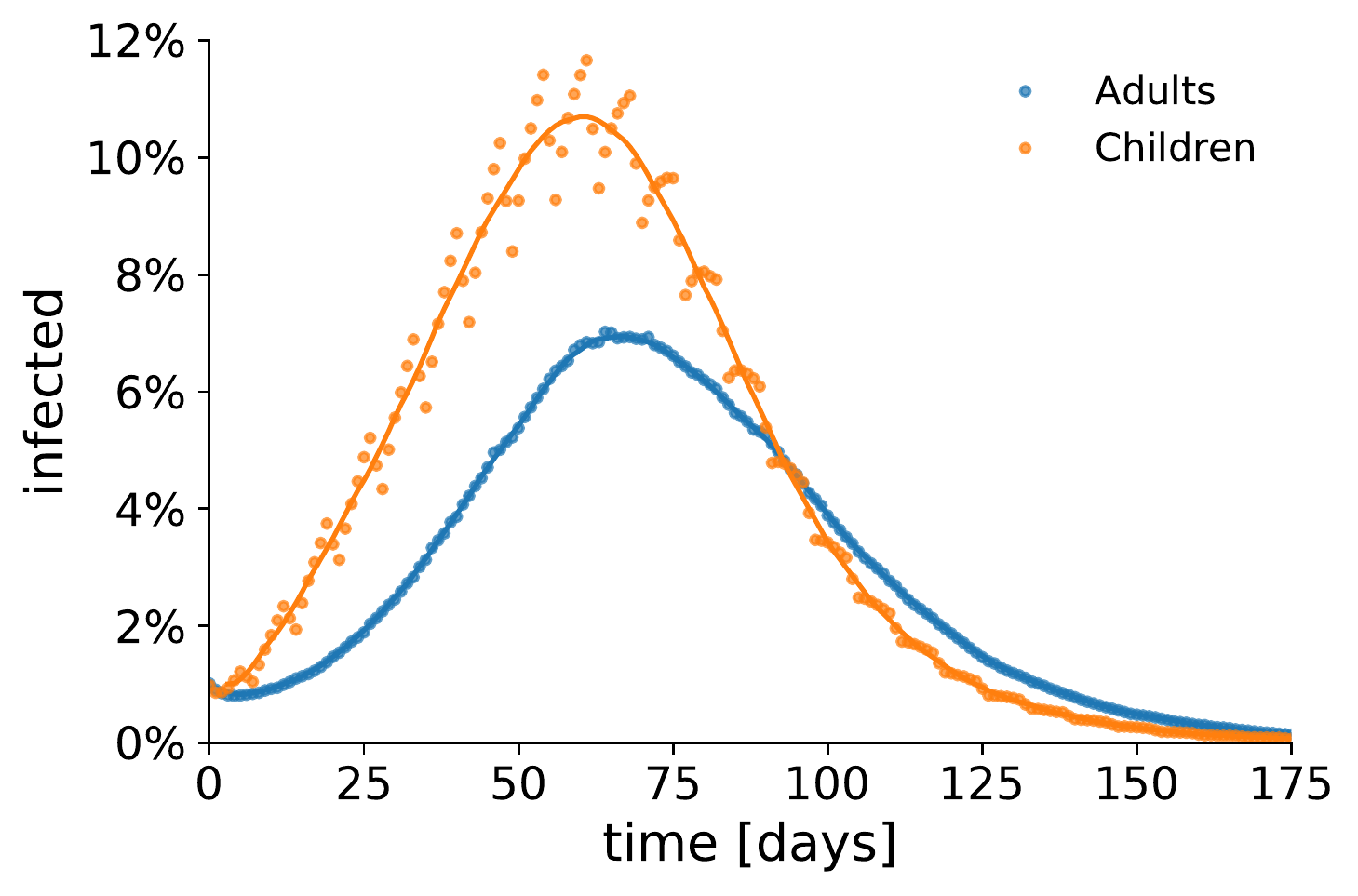}}
    \subfloat[Different vaccination strategies]{ \label{fig:vax}\includegraphics[width=7cm]{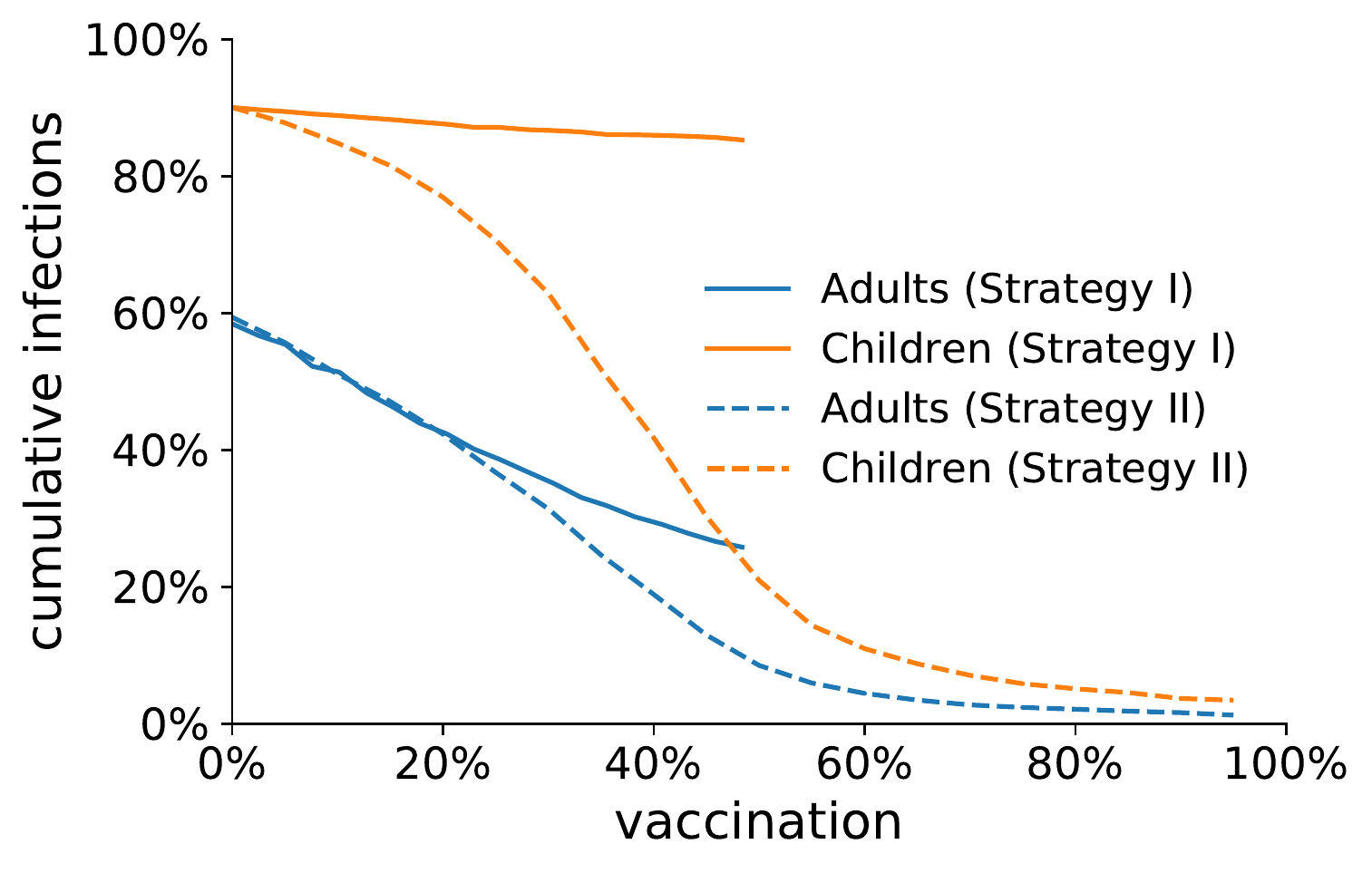}}
\caption{Role of children in the spreading of COVID-19. In (a), we show the temporal evolution of the fraction of infections among the adults (blue dots) and children (orange dots) in a representative simulation, in the absence of vaccination $V=0$. The solid curves illustrate the $7$-day moving average of the two quantities. In (b), we show the Monte Carlo estimation (over $100$ independent simulations) fraction of infections among non-vaccinated adults (blue) and children (orange) at the d of the pandemic outbreak, as a function of the fraction of vaccinated population $V$. The solid curves refer to vaccination strategy I, in which only adults are eligible for vaccination; the dotted line refers to vaccination strategy II, in which vaccine shots are randomly assigned to the entire population. Not that in strategy I, only a portion of the population is eligible for vaccination. The parameters used in the simulations are listed in Table \ref{tab:parameters}, $q_a=0.25$, and $q_c=0.12$.}
   
\end{figure}

Figure~\ref{fig:inf} reports the temporal evolution of the number of infections among the children and the adults in the absence of vaccination ($V=0$). The simulation shows that the infections among children grow faster, while the infections among adults seem to follow the children's wave. This observation intuitively suggests that control of the children spreading might be crucial toward successfully flattening the epidemic curve.

To support our intuition, we design a set of experiments to investigate the difference between two vaccination strategies, and we perform a set of Monte Carlo simulations of the epidemic process for different fractions of vaccinated  population ---spanning from no vaccinations to vaccinating $95\%$ of the eligible population, which is consistent with the maximum efficacy of the available vaccine~\cite{Polack2020}. Specifically, we compare strategy I, in which only adults are vaccinated, and strategy II, in which all the population is eligible for vaccination. In Fig.~\ref{fig:vax}, we show that strategy I has a weak impact on the epidemics, as the cumulative number of infections among children is marginally reduced, even when the entire adult population is vaccinated. Strategy II, instead, drastically impacts the epidemic spreading, yielding herd immunity when about $50-60\%$ of the population is vaccinated.

Unfortunately, the above-mentioned strategy, although attractive, could be of difficult application in the case of the COVID-19 pandemic, being children excluded from the trials of the approved vaccine so far~\cite{nyt}. Therefore, in the next sections, we will test alternative strategies to mitigate the disease, which do not entail the vaccination of children.

\subsection{Flattening the curve through testing and home-isolation policies}

\begin{figure}
    \centering
    \subfloat[Different testing effort]{\label{fig:qC}\includegraphics[width=7cm]{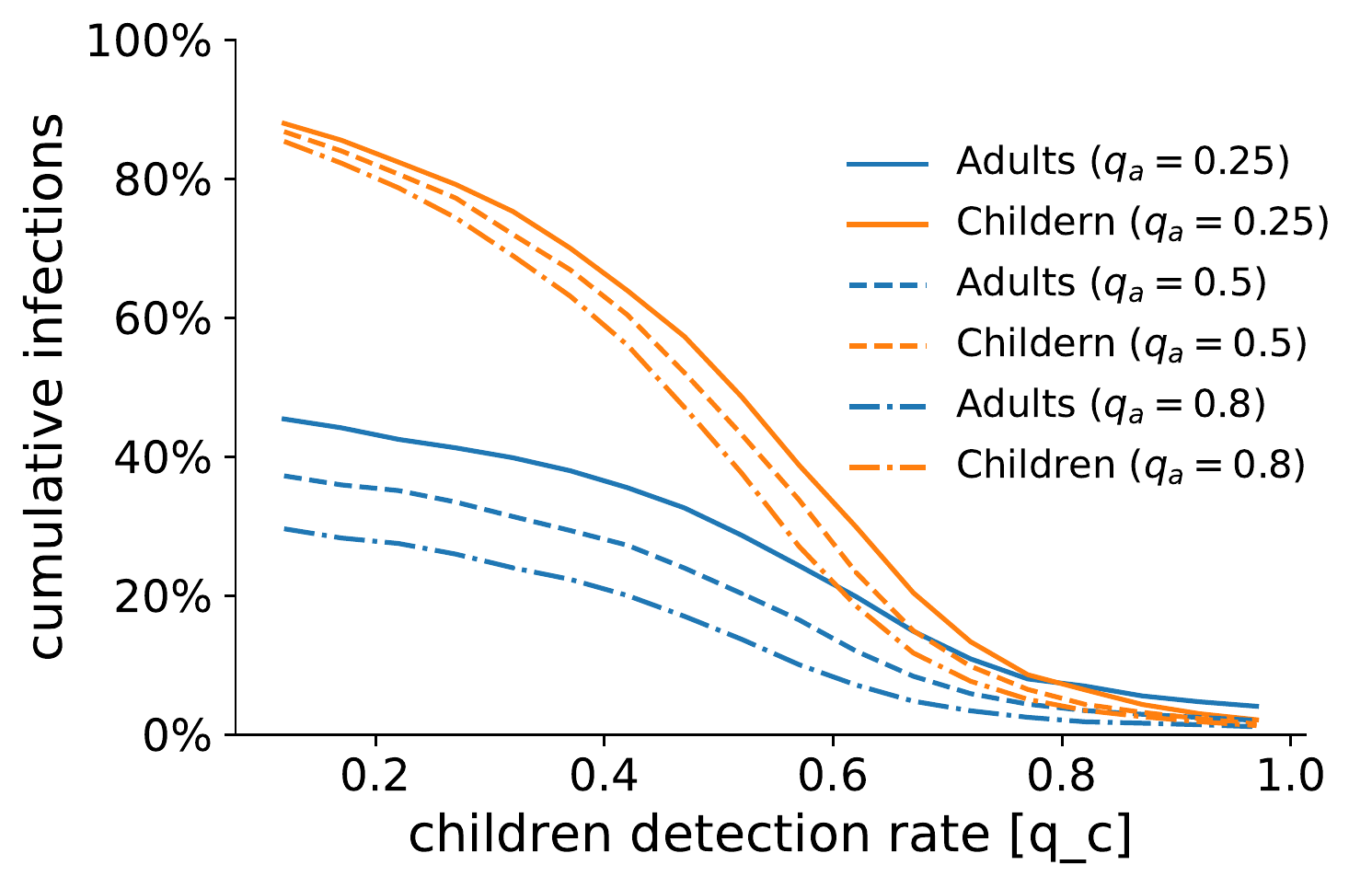}}
    \subfloat[Different home-isolation policies]{\label{fig:qCschool}\includegraphics[width=7cm]{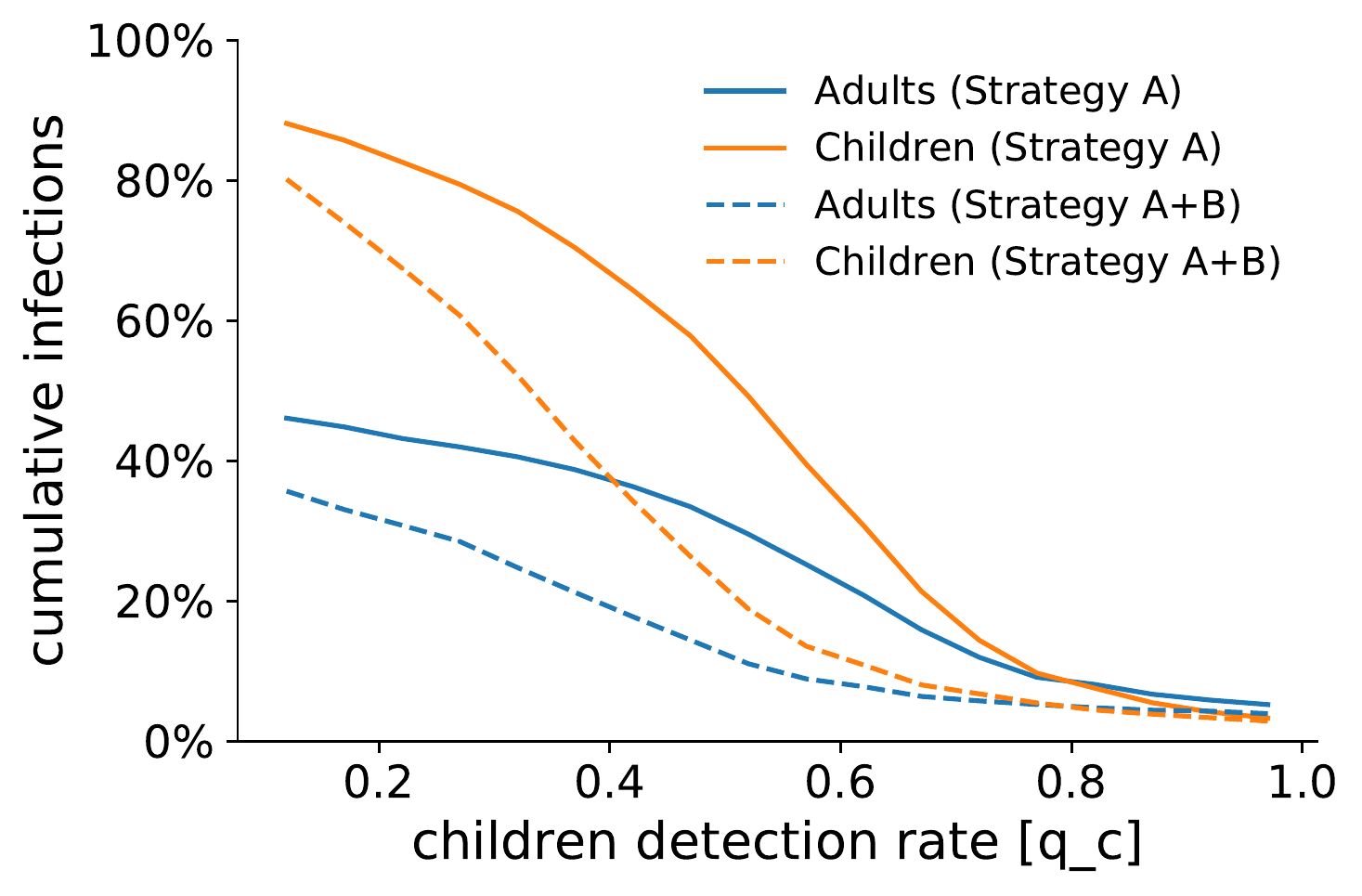}}%
    \caption{Effect of testing and different home-isolation policies. In (a), we show the Monte Carlo estimation (over $100$ independent simulations) of the cumulative fraction of infections among adults (blue) and children (orange) at the end of the pandemic outbreak as a function of the children detection rate $q_c$ and for three different values of adult detection rate, representative of low testing $q_a=0.25$ (solid), moderate testing $q_a=0.5$ (dashed), and massive testing $q_a=0.8$ (dash-dotted). In (b), we compare two home-isolation policies: family-isolation policy (A) and both the class-isolation policy (B). The Monte Carlo estimation (over $100$ independent simulations) of the cumulative fraction of infections among adults (blue) and children (orange) shows that the application of both policies (dashed) sensibly outperforms utilizing only A (solid). The parameters used in the simulations are listed in Table \ref{tab:parameters} and $V=0.17$. In (b), we set $q_a=0.25$.}
\end{figure}

In the second set of experiments, we investigate the possibility of flattening the epidemic curve by means of increasing the testing capacity and implementing effective home-isolation policies. In our simulation study, we thus fix the fraction of the vaccinated population at $17\%$ of the entire population ($V=0.17$), selected uniformly at random among the adults according to vaccination strategy I. Then, we first explore different testing strategies by estimating the cumulative number of infections for different values of $q_a\in\{0.25,0.5,0.8\}$ and $q_c\in[0.12,1]$. The results of our analysis, illustrated in Fig.~\ref{fig:qC}, suggest that, even when a non-negligible fraction of the population is already vaccinated, massive outbreaks are still possible. Moreover, we observe that the ability to detect most of the infections among adults is not sufficient to flatten the epidemic curve, as only a mild decrease in the number of cases is observed. On the contrary, massive screening campaigns among children seem to be effective in reducing the cumulative number of infections, not only among children but within the entire population. However, according to this model, effective screening policies should be able to detect at least $70\%$ of the infections among children, which may be realistically unfeasible or extremely costly, considered the large number of tests that such a practice would require to process~\cite{Pullano2020}.

To address this issue, we utilize our model to investigate whether targeted home-isolation policies for children can be enacted to reduce the testing effort needed to flatten the epidemic curve. Specifically, we consider combining the family-isolation policy A with the school-isolation policy B, described in Section~\ref{sec:home}, in which not only the family members of detected individuals have to home-isolate, but also all the classmates of children that are detected are home-isolated until their infected mate has recovered. Since we have observed that increasing the testing of adults has a mild effect, we fix $q_a=0.25$. The results are illustrated in Fig.~\ref{fig:qCschool}. Predictably, this strategy has a moderate impact for low values of $q_c$, since only a small fraction of the infected children could be detected, and thus few classes are home-isolated. As the children detection rate $q_c$ increases, the fraction of infections (among children and adults) abruptly drops down, and a detection rate of $50\%$ seems to be sufficient to guarantee successful mitigation of the outbreak. This result compares favorably to our previous findings, since it seems that disease mitigation can be achieved with more realistic testing practices.

\subsection{Targeted vaccination campaign}

\begin{figure}
    \centering
       \subfloat[Vaccination strategy I]
    {\label{fig:random}\includegraphics[width=7cm]{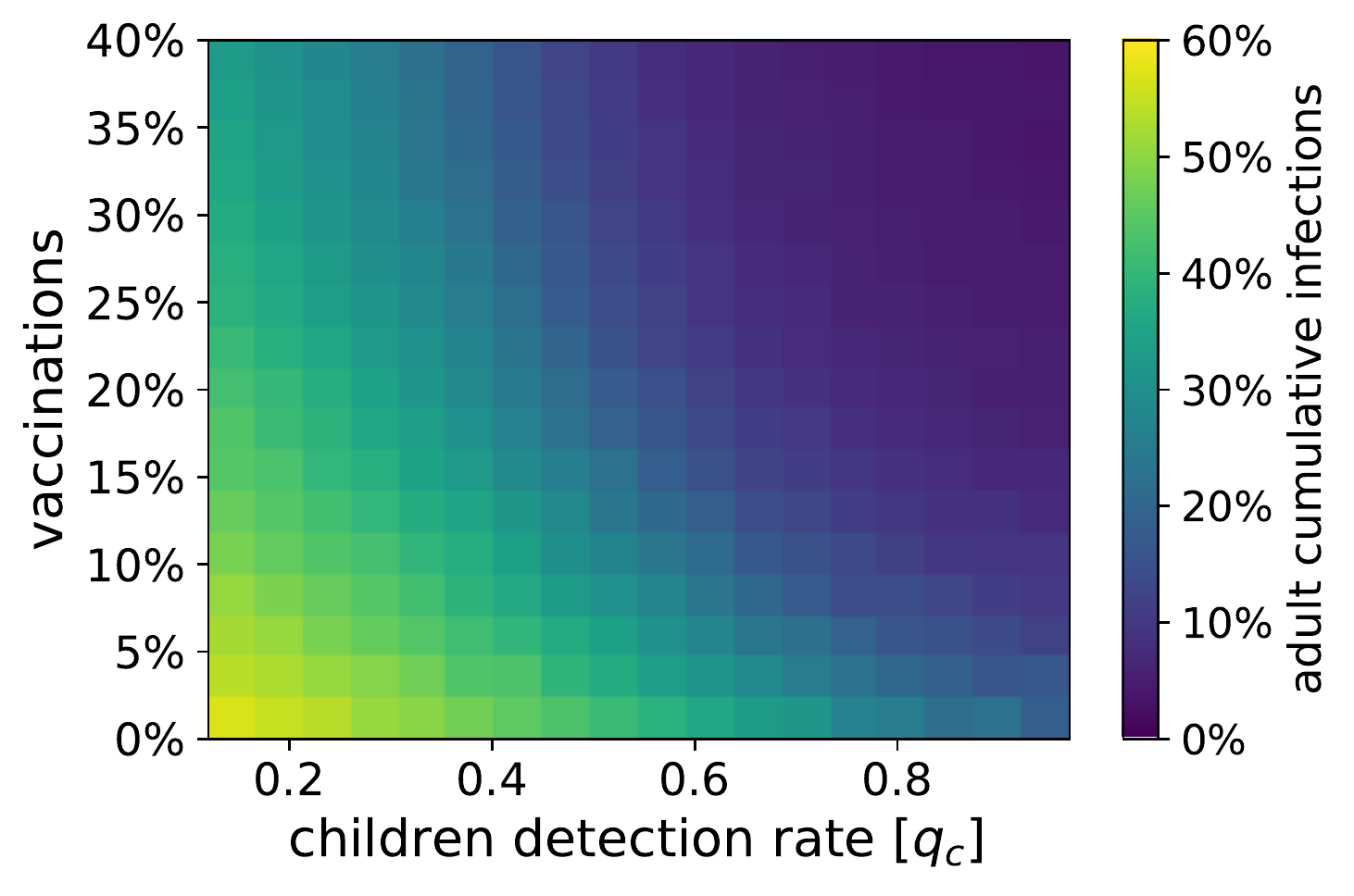}}
     \subfloat[Vaccination strategy III]{\label{fig:priority}
\includegraphics[width=7cm]{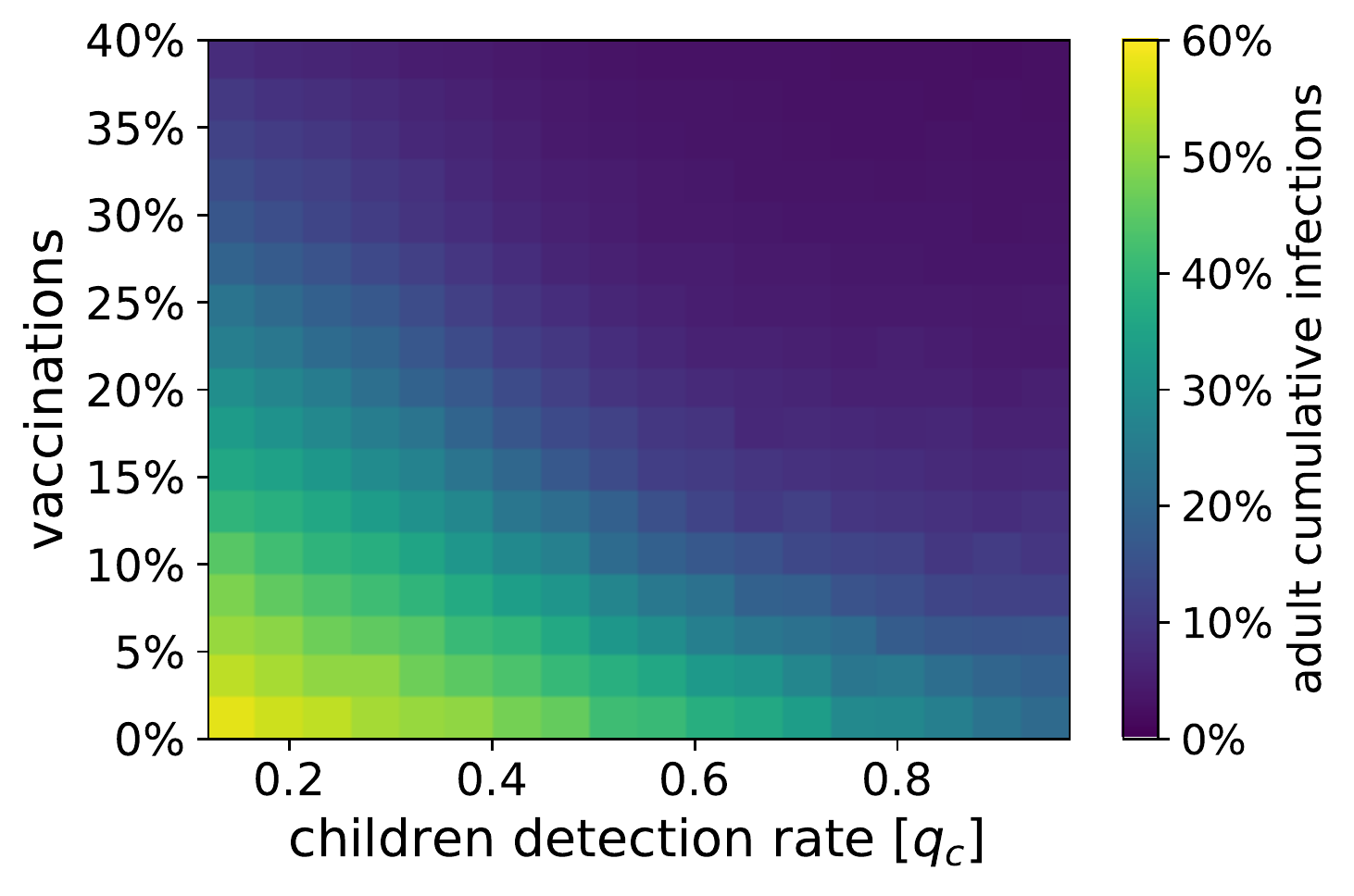}}\\
    \subfloat[Strategy comparison]
    {\label{fig:compare}\includegraphics[width=7cm]{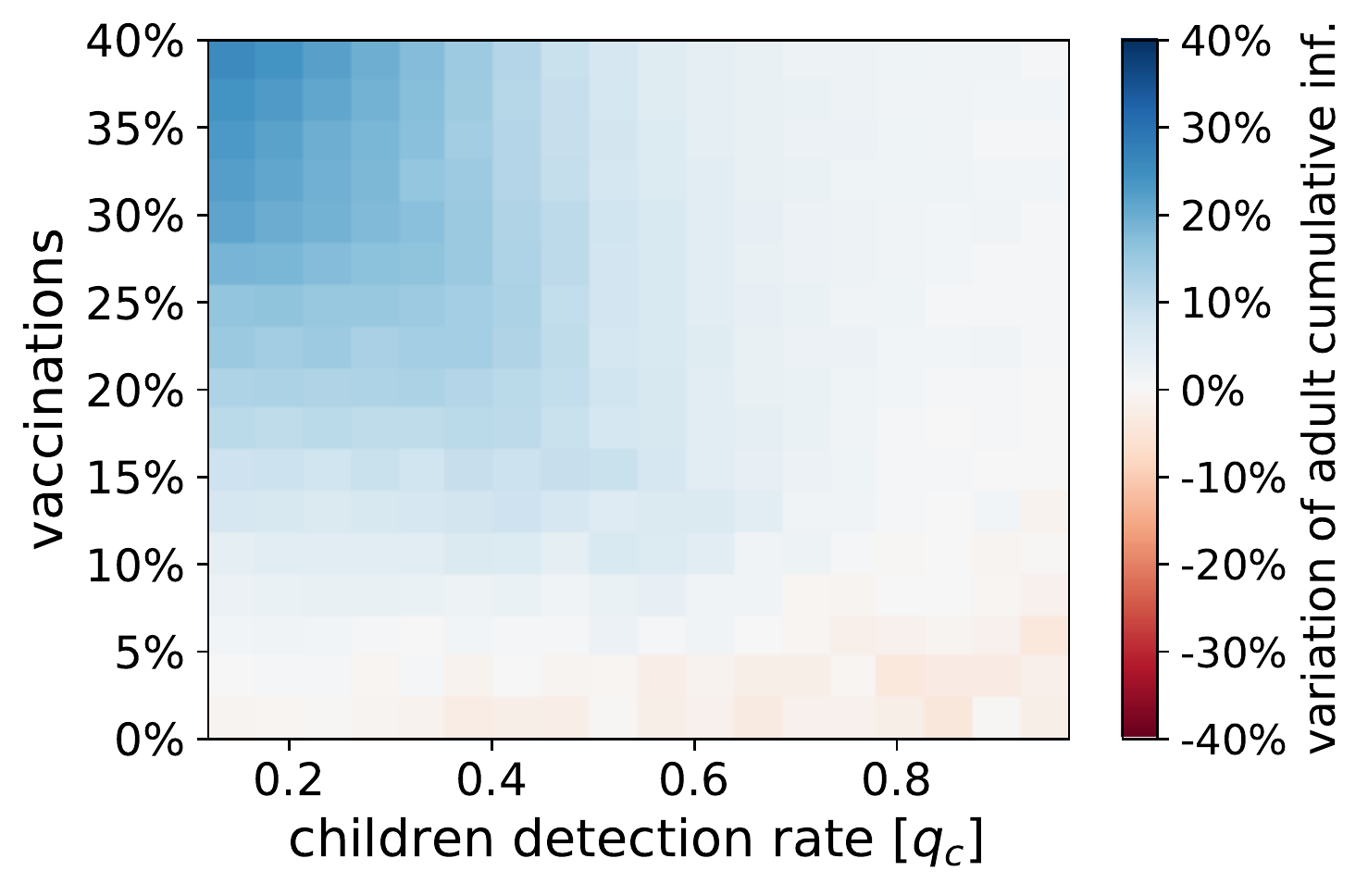}}
    \caption{Comparison between different vaccination strategies. In (a) and (b), we show the Monte Carlo estimation (over $100$ independent simulations) of the cumulative number of infections among adults at the end of the epidemic outbreak, for different values of the fraction of eligible individual vaccinated ($V$) and different children testing rate ($q_c$). In (a), we adopt vaccination strategy I, in which the vaccinated individuals are selected uniformly at random among the adult population. In (b), we adopt vaccination strategy III, in which the vaccination of adults belonging to large families is prioritized. In (c), the two strategies are compared, showing the variation in the cumulative numbers of infections between the two strategies. The parameters used in the simulations are listed in Table \ref{tab:parameters} and  $q_a=0.25$. }
    
    \end{figure}

Finally, we hypothesize that the latter strategy can be further improved by prioritizing the vaccination of families with a large number of sons. To test this hypothesis, we compare the outcome of the uniform adult vaccination strategy I, with the targeted vaccination strategy III, in which vaccinations are performed on adults, prioritizing members of large families. It is worth stressing that this approach is way more simple than prioritizing vaccination of the most active individuals based on their social activity pattern (as proposed in ~\cite{Liu2014}), being the size of each family a non-ambiguous number, easily accessible for public health authorities and regulators.  
In Fig.~\ref{fig:priority}, we show a heat-map of the cumulative number of infections among adults at the end of the outbreak under the two vaccination strategies, for different values of the children detection rate $q_c$ and of the fraction of population vaccinated $V$. For both vaccination strategies, the figure clearly depicts a trade-off between children testing and vaccinations. In plain words, when a small fraction of the adult population is vaccinated, a massive testing campaign on children is necessary to control the spreading; then, when the vaccine has reached a sufficiently large portion of the adult population, the effort placed in children testing can be reduced. 

When comparing the two vaccination strategies, we observe that, in many scenarios, strategy III (namely, the targeted vaccination) outperforms strategy I (the uniform one), as can be observed in Fig.~\ref{fig:compare}. In particular, while in the very early stages of the vaccination campaign strategy I may provide a very small benefit (in particular in the presence of massive testing campaigns), strategy III becomes more efficient when about $5\%$ of the population is vaccinated, and becomes more and more advantageous as the number of vaccinated individuals increases. For instance, when $25\%$ of the population is vaccinated, the targeted vaccination strategy reduces the infections up to $16\%$, when no testing campaigns are implemented. Interestingly, the vaccination strategy that prioritizes large families seems to yield herd immunity with $40\%$ of the population vaccinated, whereas a similar result with the uniform vaccination strategy requires a much higher number of vaccination, or a massive (and unrealistic) children testing ($q_C>0.6$).

\section{Discussion and conclusions}\label{sec:conclusions}

In this paper, we have proposed a stochastic multi-layer network model for the spreading of epidemic diseases, specifically tailored to study the current phases of the COVID-19 pandemic. In particular, the proposed model incorporates three layers of social contacts, to account for interactions between family members, within schools, and casual interactions between adults. The model is specifically oriented toward studying the role of children in the spread of COVID-19 during the ongoing vaccination phases. In these phases, NPIs are still implemented (reducing thus the number of contagions in other locations), while schools are open for in-person education in many countries. Our hypothesis is that schools play a key role in the spreading of the disease, and thus implementing intervention policies oriented to prevent contagions in schools ---such as targeted home-isolation strategies and testing campaigns--- is crucial to flatten the epidemic curve. To test our hypothesis, we put forward a campaign of Monte Carlo numerical simulations in which we have investigated the effectiveness of different home-isolation and testing policies for infected children, their families, and their schoolmates, and we have tested different vaccination strategies.

Our findings have confirmed our hypothesis. Specifically, they have provided some insights into the role of schools in the spreading of COVID-19 and into the effectiveness of different NPIs. First, our model, calibrated on COVID-19 epidemiological parameters and demographic data, has allowed us to disentangle the role of children and adults in the spreading process. Specifically, we have shown that during the current phase, contagions in schools are one of the main drivers of the epidemics, and intervention policies aiming at reducing the transmissions between children are thus key to mitigate the spread. Considered the impossibility of vaccinating children in the next few months~\cite{nyt}, the only way to control contagions in schools is by means of NPIs, such as the implementation of targeted temporary online education and testing campaigns. Our second main result suggests that a massive testing campaign of children may be effective in flattening the epidemic curve, in particular, when testing is combined with the implementation of temporary online education for the classes in which infectious children are detected. With the implementation of such an home-isolation practice, we have shown that a reasonable testing effort (able to detect about $50\%$ of the infected children) is sufficient to keep the pandemic under control. Finally, we have tested the effectiveness of a targeted vaccination policy for the next phases of the vaccination campaign, that is, after the vaccination of high-risk individuals. In the proposed strategy, the vaccination of adults in large families is prioritized. The goal of such a policy is to cut the bridges between classes associated with large families ---in which the many children are going to different classes--- which can be at the basis of those super-spreading events that are typical of the inception of COVID-19 outbreaks~\cite{Wong2020}. The results of our simulations have confirmed our intuition that the proposed vaccination strategy might be beneficial to reduce the number of infections, favoring the faster reaching of herd immunity.

In this work, we have focused on the role of schools in the spreading of COVID-19, highlighting how some intervention policies and vaccination strategies can be beneficial for mitigating the epidemics. Our findings can thus be relevant to help inform public health authorities in their decisions. However, we would like to stress that our study has some limitations. Importantly, our model is implemented assuming that the country is in partial lockdown, that is, assuming that most of the adults can work from home or in safe environments and that most of the non-essential activities are canceled (for instance, sports and cultural activities, large gatherings). Even though, as of February 2021, many countries are undergoing a (partial or total) lockdown due to the ongoing second and third waves of the COVID-19 pandemic,  addressing similar issues during the future phases of the pandemic may require to consider scenarios with the relaxation of the current NPIs. Thank the flexibility of our model, further sources of social interactions ---and, thus, of potential contagions--- can be incorporated within our modeling framework by including additional layers in the multi-layer network to capture, for instance, social interactions in work-places, in sport and cultural centers, or between groups of friends. Such a flexibility would allow the scientific community to utilize our modeling framework and adapt it to the analysis of future issues related to the control of COVID-19 and to increase preparedness for future epidemic outbreaks. 

\section*{Acknowledgment}

This work was performed using HPC resources from the ``M\'esocentre'' computing center of CentraleSup\'elec and Ecole Normale Sup\'erieure Paris-Saclay supported by CNRS and R\'egion \^{I}le-de-France. The work by L. Zino is partially supported by the European Research Council (ERC-CoG-771687).

\end{document}